\input{psfig.sty}
\documentclass[12pt]{article}
\usepackage{graphicx}
\newfont{\feff}{cmti10}
\begin{document}

\title{Mean- Field Approximation and Extended Self-Similarity 
in Turbulence}

\author{ Victor Yakhot\\
Institute for Advanced Study\\
Princeton, NJ 08540\\
and\\
Department of Aerospace and Mechanical Engineering\\
Boston University,
Boston, MA 02215 }

\maketitle

${\bf Abstract}.$
\noindent Recent experimental discovery of extended self-similarity (ESS)
was one 
of the most interesting developments, 
enabling  precise determination of 
the scaling exponents of fully developed turbulence. Here we show that the 
ESS is consistent with the Navier-Stokes equations, 
provided the pressure -gradient 
contributions are expressed in terms of velocity 
differences in  the mean field approximation (
Yakhot, Phys.Rev. E{\bf 63}, 026307, 
(2001)). A sufficient  condition for extended self-similarity 
in a general dynamical system
is derived.

\newpage

Scaling relations for velocity  structure functions in isotropic
and homogeneous turbulence are   defined as :

\begin{equation}
S_{n,m}=
<(u({\bf x+r},t)-u({\bf x},t))^{n}(v({\bf x+r},t)-v({\bf x},t))^{m}>=
c_{n,m}({\cal E}r)^{\frac{n+m}{3}}(\frac{r}{L_{f}})^{\xi_{nm}-\frac{n+m}{3}}\phi_{n,m}(\frac{r}{L_{f}},\frac{r}{\eta})
\end{equation}

\noindent where  $u$ and $v$ are 
components of  velocity field parallel and perpendicular to 
the displacement vector ${\bf r}$, respectively. 
The universality assumption implies that the coefficients $c_{n,m}=O(1)$,
independent of the Reynolds number (dissipation scale 
$\eta\approx L_{f}Re^{-\frac{3}{4}}$). The dissipation rate ${\cal E}=
\overline{(\partial_{i}u_{j})^{2}}=const=O(1)$ is equal to the power of 
external kinetic energy pumping. The shape of the structure functions (1)
is an assumption , not following any  rigorous theory.  In 
the inertial range ($\frac{r}{L_{f}}\rightarrow 0$ , 
$\frac{r}{\eta}\rightarrow \infty$) the  
scaling functions $\phi_{nm}(r)\rightarrow a_{n,m}=const$, 
 independent of the displacement $r$. 

Both physical and numerical experiments show that the functions 
$\phi_{n,m}$ start  deviating from the constant inertial range values at $ r/\eta\approx 10$.
Since one
 does not have theoretical 
expressions for $\phi_{n,m}$, accurate measurements
of  exponents $\xi_{n,m}$ 
in a fully developed turbulent flow requires  an
extremely wide  range  of variation  of the displacement   
$r$ which is  possible only  if the 
Reynolds number of a flow is huge. 
This problem is even more severe 
for  numerical simulations of turbulence,  where usually the wide inertial 
range is difficult  to generate. 

\noindent It has been shown in  a 
remarkable paper by Benzi et al [1]
that even in the medium (quite low, actually) Reynolds number 
flows, where (1) is hard to observe
, the following 
relation (ESS) holds:

\begin{equation}
S_{n,0}(r)=C_{n0}S_{m,0}(r)^{\beta(nm)}
\end{equation}

\noindent where $\beta(nm)=\frac{\xi_{n}}{\xi_{m}}$. It is clear from (1)
that if $c_{n,m}$ are reynolds number independent, then the coefficients
$C_{n,m}$ in (2) do not depend on the dissipation scale 
$\eta$ (Reynolds number).  
Since  the range of validity of expression (2) is much 
wider 
than that of
(1),  accurate determination  of exponents $\beta(nm)$ enables one 
to evaluate the exponents $\xi_{nm}$ even in the not-too-high Reynolds 
number flows. Comparison of the exponents calculated this way with 
those measured in extremely high  Re flows ($\beta(nm)\approx \xi_{n,0}/\xi_{m,0}$) was usually  extremely good [3]. 
Since its discovery the relation (2) evolved into a major tool
for experimental and numerical 
determination of the exponents $\xi_{nm}$ [1]-[5]. The definition (2) 
was introduced and tested in Ref. [2]. Since $S_{3,0}\propto r$ in the inertial range,
it is is  typically used in  application of the ESS (2) 
for analysis of experimental  data. 
It is shown below that extended self-similarity (2) can be derived 
self-consistently 
from the Navier-Stokes equations.

\noindent 
It was shown that in a statistically isotropic, homogeneous 
 and incompressible flow 
governed  by the Navier-Stokes equations the following equation can be
 rigorously derived in the limit $r/L_{f}\rightarrow 0$ 
where the forcing function can be neglected [6] :

\begin{equation}
\frac{\partial S_{2n,0}}{\partial r}+\frac{d-1}{r}S_{2n,0}-
\frac{(d-1)(2n-1)}{r}S_{2n-2,2}=
-(2n-1){\cal P}_{x, 2n-2}+(2n-1)\nu D_{u,2n-2}
\end{equation}

\noindent 
where

\begin{equation}
{\cal P}_{x,2n-2}=\overline{(p_{x'}(x')-p_{x}(x))(\Delta u)^{2n-2}}
\end{equation}

\noindent and

\begin{equation}
D_{u,n}=\overline{(\nabla^{2} u(x')-\nabla^{2} u(x))(\Delta u)^{n}}
\end{equation}

\noindent These relations are exact even in the low- 
Reynolds- number statistically isotropic and homogeneous flows in 
the range $r/L_{f}\rightarrow 0$. It is important that $D_{u,2n}(r)=O(1)$ and 
thus, $\nu D_{u,2n}\rightarrow 0$ as $\nu\rightarrow 0$ in the inertial range.
On the other hand, due to the dissipation anomaly, 
 $\nu D_{u,2n+1}$ is finite  in the inertial range.
To prove the former  statement,  we consider:

\begin{equation}
(2n-1)\nu D_{u,2n-2}=-(2n-1)(2n-2)\overline{({\cal E}_{u}(2)+{\cal E}_{u}(1))(\Delta u)^{2n-3}}+
\nu \partial_{r}^{2} S_{2n-1,0}(r)
\end{equation}

\noindent where ${\cal E}_{u}=\nu\overline{(\partial u)^{2}}$. 
The second term in (6) disappears in the inertial range 
in the limit $\nu\rightarrow 0$. To estimate the first contribution,  we write
neglecting the subscript $u$:

\begin{equation}
\overline{({\cal E}(2)+{\cal E}(1))(\Delta u)^{2n-3}}\leq \sqrt{\overline{({\cal E}(1)+{\cal E}(2))^{2}}}S_{\frac{4n-6}{2}}(r)
\end{equation}

\noindent Since  in the inertial range 
($r\rightarrow 0$), 
 $\overline{({\cal E}(1)+{\cal E}(2))^{2}}\propto r^{-\mu}$
with $\mu\approx 0.2$, 
 this term is negligibly small compared to the 
$O(S_{2n,0}/r)$ contributions to (3) for not too small moment number $n$, 
provided $\xi_{2n,0}$ ``bends''  strong enough with $n$. This is 
definitely true  on the expression 

\begin{equation}
\xi_{2n,0}=\frac{1+3\beta}{3(1+2n \beta)}2n
\end{equation}

\noindent derived in [6]. In the inertial range the dissipation 
contributions to (3) can be neglected. This does not mean that the even-order 
structure functions are not affected by the dissipation processes.
The equation (3) is not closed and as a result the even -order
 moments are 
coupled to the dissipation contributions appearing in the equations for
 the odd-order moments. This will be discussed below.

\noindent The equation  (3) is a direct consequence of
the Navier-Stokes equations. We will  show in what follows that the 
ESS is consistent with (3). Let us,  in accord with ESS (2),  assume
that $S_{2n}=S_{2n}(S_{2m})$ where $m$ is an arbitrary number. 
This assumption  is non-trivial since, in principle, the moment $S_{2n}$
 can also depend  on the displacement $r$ and  
dissipation scale $\eta$ (Reynolds number).
Substituting this into (3) gives:

\begin{equation}
\frac{\partial S_{2n,0}}{\partial S_{2m,0}}=\frac{(d-1)S_{2n,0}-
(d-1)(2n-1)S_{2n-2,2}+(2n-1)r{\cal P}_{x,2n-2}-(2n-1)\nu r D_{u,2n-2}}{(d-1)S_{2m,0}-
(d-1)(2m-1)S_{2m-2,2}+(2m-1)r{\cal P}_{x,2m-2}-(2m-1)\nu r D_{u,2m-2}}
\end{equation}

\noindent The relation (2) holds if the right side of (9) is equal to 
$\frac{\xi_{2n,0}S_{2n,0}}{\xi_{2m,0}S_{2m,0}}$. 
Again, the relations (9) are exact everywhere as long as 
$r/L_{f}\rightarrow 0$. 
The mean field approximation, introduced  in [6],  is a 
statement that 
the pressure-gradient 
difference is expressible in term of a quadratic form of velocity differences.
Since $<\Delta p_{y}(\Delta u)^{2}>= 
<\Delta p_{y}(\Delta v)^{2}>        =0$,
we are left with:

\begin{equation}
\Delta p_{x}=\frac{a (\Delta u)^{2}+
b(\Delta v)^{2}}{r}+c\frac{d}{dr}(\Delta u)^{2}+ \cdot\cdot\cdot
\end{equation}

\noindent The coefficients  $a$, $b$ and $c$ etc are chosen
so that $\overline{\Delta p_{x}}=\overline{\Delta p_{x} \Delta u}=
\overline{\Delta p_{x} \Delta v}=0$. We also have 
 (see [6]):

\begin{equation}
\Delta p_{y}\propto \Delta u \Delta v/r
\end{equation}

\noindent The equations (3)  (9)-(11)  are not closed since we not have 
the relations coupling $S_{2n,0}$ with $S_{2n-2,2}$. We know that in 
the dissipation range, $r/\eta\rightarrow 0$  the functions 
  $S_{2n,0}\propto S_{2n-2,2}$ and 
$\xi_{2n,0}=2n$, while in the inertial range the correlation 
functions are characterized by the non-trivial exponents (1). 
In principle, based on [6], we can easily write equations for $S_{2n-2,2}$.
However, they involve the correlation functions $S_{2n-4,4}$ etc.

\noindent Now we would like to ask the central  question: consider
a relatively low Reynolds number flow , so that the dissipation contributions
to (3) cannot be neglected and the functions $\phi_{2n,0}(0,\frac{r}{\eta})$ 
vary with the displacement $r$. What is the structure of the theory
 preserving (2) but strongly violating the inertial range scaling 
$S_{2n,0}\propto r^{\xi_{2n,0}}$? At the top of the dissipation range
$r/\eta \approx 1-10$ the scaling functions, violating the inertial 
range scaling are not small (see (1)). For $2m=2$
 the equation (9) simplifies:

\begin{equation}
\frac{\partial S_{2n,0}}{\partial S_{2,0}}=\frac{(d-1)S_{2n,0}-
(d-1)(2n-1)S_{2n-2,2}+(2n-1)r{\cal P}_{x,2n-2}-(2n-1)\nu r D_{u,2n-2}}
{(d-1)(S_{2,0}-S_{0,2})}
\end{equation}

\noindent where
 in 
incompressible, isotropic and homogeneous turbulence  
$(d-1)(S_{0,2}-S_{2,0})=-r\frac{d S_{2,0}}{dr}$ (see (3)). 
Both dissipation and pressure contributions do not appear 
in the denominator of (12).  
The form of relation (12) tells us 
that the ESS  
$S_{2n,0}={\cal C}_{2n,2}(S_{2,0})^{\frac{\xi_{2n,0}}{\xi_{2,0}}}$,
 exact in the inertial range if the relations (1) are valid,  is 
possible only 
in an interval where 
the numerator  in (12) is equal to $-r\frac{d S_{2n,0}}{d r}$. 

Substituting this  into (12) and using the scaling form (1) gives:

\begin{equation}
\frac{\partial S_{2n,0}}{\partial S_{2,0}}=
\frac{\frac{d S_{2n,0}}{d r}}{\frac{d S_{2,0}}{d r}}=
\frac{\xi_{2n,0}S_{2n,0}}{\xi_{2,0}S_{2,0}}
\frac{1+\frac{x}{\xi_{2n,0}\phi_{2n,0}(x)}\frac{d\phi_{2n,0}(x)}{dx}}
{1+\frac{x}{\xi_{2n,0}\phi_{2,0}(x)}\frac{d\phi_{2,0}}{dr}}
\end{equation}

\noindent By assumption,  $S_{2n}=S_{2n}(S_{2})$,  subject to 
``boundary condition'' $S_{2n,0}=C_{2n,2}S_{2}^{\frac{\xi_{2n,0}}{\xi_{2,0}}}$ as
$x\rightarrow \infty$. 
Solution to (13), satisfying these  constraints, is: 
$\phi_{2n,0}\propto \phi_{2,0}^{\frac{\xi_{2n,0}}{\xi_{2,0}}}$. 
Indeed, substituting this into the second equation 
(13), we are left with the differential equation, equivalent to the ESS (2)
with the Reynolds-number-independent coefficient 
$C_{2n,2}$. One can see that the 
ESS is the only universal 
solution to the equation (13), not involving any dependence on the Reynolds number ($\eta$).

Since (3) and (9)
are  a direct consequence 
of the equations of motion for  velocity field, we conclude that the
the ESS with
non-trivial scaling exponents is 
consistent with the Navier-Stokes equations as long as the scaling 
assumption (1) is valid.

The function  $\phi_{2,0}(x)$ can be readily self-consistently 
found from the well-known 
differential equation:

\begin{equation}
S_{3,0}=-0.8 r+6\nu\frac{d S_{2,0}}{d r}
\end{equation}

\noindent The inertial range calculations [6] and both numerical and 
physical experiments [3]  give $\xi_{2,0}\approx 0.7$ and 
$\phi_{2,0}(x)\approx a_{2,0}\approx 2.0$
(Kolmogorov constant $C_{K}\approx 1.6$). 
Substituting the ESS expression 
$S_{3,0}^{\xi_{2,0}}\propto S_{2,0}$ into (14) gives:

\begin{equation}
\frac{6 d\phi_{2,0}}{dx}=(0.8~-~0.3\phi_{2,0}^{\frac{1}{\xi_{2,0}}})
x^{1-\xi_{2}}-6\xi_{2}\phi_{2,0}/x
\end{equation}

\noindent where by definition of the dissipation scale 
$\nu\eta^{-2+\xi_{2}}=1$ and ${\cal E}=1$. Solution 
to this equation gives $\phi_{2,0}(x)$,  gently approaching 
$a_{2,0}\approx 2$
as $x\rightarrow \infty$. Noticible deviations from this 
constant value  start at $x\approx 30-50$ (at $x=20$ the function 
$\phi_{2,0}\approx 1.65-1.7$). This  equation was derived by 
Benzi et. al. (Ref.[2]).

In the inertial range, where the dissipation 
contributions to (12) are negligible and where the ESS is sinonimous to the 
power laws,  the numerator of (12) is equal to $-r\frac{d S_{2n,0}}{dr}$ and
the mean field approximation as  exact as the power laws themselves. 
It is interesting that even if  $\xi_{2n,0}\neq S_{2n-2,2}$, the power
 laws solutions of (3) are still
possible: in principle the mixed moments  $S_{2n-2,2}$ can be
 cancelled by
the corresponding pressure-gradients  contributions to (10).

To conclude:  It follows directly from the Navier-Stokes equations that
if  the inertial range scaling exists,  then:

\begin{equation}
(d-1)S_{2n,0}-
(d-1)(2n-1)S_{2n-2,2}+(2n-1)r{\cal P}_{x,2n-2}=-r\frac{d S_{2n,0}}{d r}
\end{equation}

\noindent This expression means that the inertial range pressure contribution 
to this equation must be $O(S_{2n,0})$ or $O(S_{2n-2,2})$. This proves the 
mean-field approximation (10). 

It can be shown that in the interval $x>1$, the direct 
dissipation contribution
to (12) is small. Then,  the mean -field approximation justifies the 
assumption $S_{2m,0}=S_{2m,0}(S_{2,0})$.  This leads to the ESS.

However, the general statement, not related to a particular dynamical 
system,  
 can be made: 1.~if the scaling relation (1) with the O(1) coefficients
$c_{n,m}$ is valid;~ 2.~ if $S_{n,0}=S_{n,0}(S_{m,0})$ 
 is independent on $r$ and $\eta$,
then $S_{n,0}=C_{n,m}S_{m,0}^{\frac{xi_{n,0}}{\xi_{m,0}}}$. This relation 
means that there exist only one dominating (dynamically relevant) 
scaling function $\phi_{i,j}$
and all others can calculated in a simple way. 

Now we can discuss the cases where the ESS is violated. 
In a strongly sheared wall flow one can introduce two Reynolds numbers.
The first one is $Re=\overline{U}L/\nu$ where $L$ is the 
width of the channel (boundary layer, etc) and $\overline{U}$ is a characteristic (mean) velocity. The second one ($Re=u_{*}L/\nu$ )
is based on the friction velocity $
u_{*}^{2}=-\nu\frac{\partial U}{\partial y}|_{wall}$. The dissipation rate
${\cal E}=O(\frac{U^{3}}{L}Re_{*}^{4}/Re)$ is a weak function of
the Reynolds number (dissipation ) scale. Thus, all structure functions, 
even if they can be written in a form (1), must involve the 
Re-dependent proportionality coefficients. This violates the assumptions
leading to the ESS (2). Far enough from the wall, 
where ${\cal E}\approx \overline{U}^{3}/L$ with the 
Re-independent proportionality coefficient one can 
expect the ESS to be valid.  

In some sheared flows $1/L_{s}=\frac{\partial u}{\partial y}/u=O(1)$,
the scaling functions also depend on $y/L_{s}\approx 1$. In these regions the 
$y/L_{s}$ cannot be neglected and the simple derivation of the ESS (13) 
breaks down. The best example, illustraiting this point, is the 
Kolmogorov flow driven by the forcing function ${\bf f}=(0,cos L_{s}x)$. 
There we expect the ESS to hold in the vicinities of zeros 
of the local strain rate $\partial_{x} U_{y}\propto sin(L_{s}x)$ and
break down near local maxima (minima) of the strain rate.  These 
conclusions agree with the experimental findings [9], [10].

\noindent I am grateful to S.Kurien, K. R. Sreenivasan , A. Polyakov 
and M. Vergassola
for their help and comments .

\section{References}
1.R. Benzi, S.Ciliberto,R.Tripiccione, C.Baudet, S.Succi,
Phys.Rev.E{\bf 48}, R29 (1993)\\
2. R. Benzi,S.Ciliberto, G.Ruiz Chavarria, Physica D {\bf 80}, 385 (1995)\\
3. K.R.Sreenivasan and B. Dhruva, Prog.Theor.Phys. {\bf 130},103 (1998)\\
4. S. Kurien and K.R. Sreenivasan, Phys. Rev. E, in press (2001)\\
5. N.Cao, S. Chen and Z.She, Phys.Rev.Lett. {\bf 86}, 3775 (2001)\\
6. V. Yakhot , Phys.Rev.E {\bf 63}, 026307 (2001)\\
7. A.N. Kolmogorov, 1941\\
8. B. Dhruva and K.R. Sreenivasan, private communication\\
9. G. Stolovitzky and K.R. Sreenivassan, Phys. Rev.E,  {\bf 48}, 32 (1993)\\
10. V. Borue, private communication

\end{document}